\documentclass[11pt]{article}
\usepackage{amsmath,amssymb,latexsym,graphicx}

\begin{document}

\begin{flushright}
	KUNS-1972 \\
	YITP-05-22 \\
	OIQP-05-05 \\
	hep-th/0505237
\end{flushright}

\vspace{1cm}

\begin{center}
{\Large\bf Supersymmetric Relativistic Quantum Mechanics}\\
\vspace{2.5cm}
{\large Yoshinobu H}ABARA\\
\vspace{0.2cm}
{\it  Department of Physics, Graduate School of Science,}\\
{\it  Kyoto University, Kyoto 606-8502, Japan}\\
\vspace{1cm}
{\large Holger B. N}IELSEN\\
\vspace{0.2cm}
{\it  Niels Bohr Institute, University of Copenhagen,}\\
{\it  17 Blegdamsvej Copenhagen \o, Denmark}\\
\vspace{0.3cm}
and\\
\vspace{0.3cm}
{\large Masao N}INOMIYA\footnotemark \\
\vspace{0.2cm}
{\it  Yukawa Institute for Theoretical Physics,}\\
{\it  Kyoto University, Kyoto 606-8502, Japan}
\end{center}

\footnotetext{Also working at Okayama Institute for Quantum Physics, Kyoyama-cho 1-9, Okayama-city 700-0015, Japan}

\vspace{1.5cm}

\begin{abstract}
We present an attempt to formulate the supersymmetric and relativistic quantum mechanics in the sense of realizing supersymmetry on the single particle level, by utilizing the equations of motion which is equivalent to the ordinary 2nd quantization of the chiral multiplet. The matrix formulation is used to express the operators such as supersymmtry generators and fields of the chiral multiplets. We realize supersymmetry prior to filling the Dirac sea.
\end{abstract}

\vspace{2cm}

\section{Introduction}

In the supersymmetric theories there seems to exist no truely quantum mechanical and relativistic, i.e. one particle model, so that one starts from the field theories. We propose a method to construct sypersymmetric quantum mechanics. In our previous papers~\cite{hnn2}, we obtained only the off-shell representation which contains the time derivative $\partial_0$ in the operators such as the supersymmetry generators, however, it is unsatisfactory representation because the time coordinate should be treated not as a coordinate but a parameter in the formulation of the quantum mechanics of the energy conserved system. 

In this paper, we would like to employ the one particle supersymmetric theory which may be considered 1st excited states in the boson vacuum, i.e. filled Dirac seas for bosons as well as fermions in our formalism~\cite{nn,hnn,nn2,hnn4}. This theory is viewed as supersymmetric relativistic quantum mechanics of the massive chiral multiplet. We propose a prescription how to realize the supersymmetry in a single particle level. Furthermore the on-shell representation which does not contain the time derivative is presented by utilizing the equations of motion. In doing so, we employ the matrix formulation whose entries are the fields of the chiral multiplets and the operators such as differentials.

\section{Off-shell Representation}

We attempt to formulate the realization of the supersymmetric one particle system, i.e. relativistic quantum mechanics. The formulation in our 2nd qunantization method is really starting point~\cite{hnn}: The one particle system appears as the excitation of the unfilled Dirac seas for bosons and fermions.

We first present the action for the chiral multiplet in the N=1 supersymmetric theory in 4 dimensional space-time, 

\begin{align}
	S=\int d^4x \left\{ -\vec{\partial}_{\mu}A^{\ast}\vec{\partial}^{\mu}A
	-i\bar{\psi}\bar{\sigma}^{\mu}\partial_{\mu}\psi +F^{\ast}F
	+m\left( AF+A^{\ast}F^{\ast}\right)
	-\frac{m}{2}\left(\psi \psi +\bar{\psi}\bar{\psi} \right)\right\}, 
\end{align}

\noindent which is invariant under the supersymmetry transformation, 

\begin{align}
	& \delta_{\xi}A=\sqrt{2}(\xi \psi ), \nonumber \\
	& \delta_{\xi}\psi_{\alpha}
	=i\sqrt{2}(\sigma^{\mu}\bar{\xi})_{\alpha}
	\vec{\partial}_{\mu}A+\sqrt{2}\xi_{\alpha} F, \nonumber \\
	& \delta_{\xi}F=i\sqrt{2}(\bar{\xi}\bar{\sigma}^{\mu}
	\vec{\partial}_{\mu}\psi ).
\end{align}

\noindent Here, for the notation of the spinor indices, see Appendix (and also Ref.~\cite{wess} for details). The three fields $A, \psi_{\alpha}$ and $F$ obey the equations of motion: 

\begin{align}
	& \vec{\partial}_{\mu}\vec{\partial}^{\mu}A+mF^{\ast}=0, \nonumber \\
	& i(\bar{\sigma}^{\mu}\vec{\partial}_{\mu}\psi )^{\dot{\alpha}}
	+m\bar{\psi}^{\dot{\alpha}}=0, \nonumber \\
	& F+mA^{\ast}=0.
\end{align}

\noindent We would like to employ the expression of the generators and fields in terms of the matrices. For this purpose we write the chiral multiplet as the vector notation: 

\begin{align}
	\Phi^C=
	\left( \begin{array}{@{}c@{}}
	A \\ \psi_{\alpha} \\ F
	\end{array} \right).
\end{align}

\noindent Then the supersymmetry generators at the 1st quantized level in an off-shell representation on the chiral multiplet which induces the supersymmetry transformation (2) is given by 

\begin{align}
	Q_{\beta}^C=
	\left( \begin{array}{ccc}
	0 & \sqrt{2}\delta_{\beta}^{\> \alpha} & 0 \\
	0 & 0 & \sqrt{2}\epsilon_{\alpha \beta} \\
	0 & 0 & 0
	\end{array} \right), \quad 
	\bar{Q}_{\dot{\beta}}^C=
	\left( \begin{array}{ccc}
	0 & 0 & 0 \\
	-i\sqrt{2}\sigma^{\mu}_{\alpha \dot{\beta}}\vec{\partial}_{\mu} 
	& 0 & 0 \\
	0 & i\sqrt{2}\epsilon_{\dot{\beta}\dot{\delta}}
	\bar{\sigma}^{\mu \dot{\delta}\alpha}\vec{\partial}_{\mu} & 0
	\end{array} \right).
\end{align}

The Lorentz-invariant inner product, in other words, the density of probability is defined as natural one, 

\begin{align}
	& \langle \Phi |\Phi \rangle \equiv \int d^3\vec{x} \> 
	\Phi^{C\dagger}I^C\Phi^C=\int d^3\vec{x} \> 
	\left\{ \frac{i}{2}A^{\ast}\overleftrightarrow{\partial_0}A 
	-\frac{1}{2}\left(\bar{\psi}\bar{\sigma}^0\psi \right)\right\},
\end{align}

\noindent with a matrix $I^C$,

\begin{align}
	& I^C\equiv \left( \begin{array}{ccc}
	\frac{i}{2}\overleftrightarrow{\partial_0} & 0 & 0 \\
	0 & -\frac{1}{2}\bar{\sigma}^{0\dot{\alpha}\alpha} & 0 \\
	0 & 0 & 0 \\
	\end{array} \right).
\end{align}

It is easy to check that the supersymmetry algebra as follows: 

\begin{align}
	\{Q_{\beta}^C,\bar{Q}_{\dot{\beta}}^C\}\Phi^C
	& =\left( \begin{array}{c}
	-2i\sigma^{\mu}_{\beta \dot{\beta}}\vec{\partial}_{\mu}A \\
	2i\epsilon_{\alpha \beta}\epsilon_{\dot{\beta}\dot{\delta}}
	\bar{\sigma}^{\mu \dot{\delta}\gamma}\vec{\partial}_{\mu}\psi_{\gamma}
	-2i\sigma^{\mu}_{\alpha \dot{\beta}}\vec{\partial}_{\mu}\psi_{\beta} \\
	-2i\sigma^{\mu}_{\beta \dot{\beta}}\vec{\partial}_{\mu}F \\
	\end{array} \right) \nonumber \\
	& =-2i\sigma^{\mu}_{\beta \dot{\beta}}\vec{\partial}_{\mu}\Phi^C,
\end{align}

\noindent and the hermiticity condition, 

\begin{align}
	\left<Q\Phi|\Phi\right>_{\dot{\beta}}
	& = \int d^3\vec{x} \> \Phi^{C\dagger}
	\big(Q^C\big)_{\dot{\beta}}^{\dagger}I^C\Phi^C \nonumber \\
	& = -\frac{1}{\sqrt{2}}\int d^3\vec{x} \> \left\{ 
	iA\overleftrightarrow{\partial_0}\bar{\psi}_{\dot{\beta}}
	-mA(\psi \sigma^0)_{\dot{\beta}} \right\} \nonumber \\
	& = \int d^3\vec{x} \> \Phi^{C\dagger}I^C\bar{Q}_{\dot{\beta}}^C\Phi^C 
	\nonumber \\
	& =\left<\Phi|\bar{Q}\Phi\right>_{\dot{\beta}},
\end{align}

\noindent holds by utilizing the equations of motion (3), where the dagger $\dagger$ on $\Phi^C$ and $Q_{\beta}^C$ denotes the usual hermitian conjuation of the matrix. In other words, by means of the equations of motion, the identity of matrices, 

\begin{align}
	(Q_{\beta}^C)^{\dagger}I^C=I^C\bar{Q}_{\dot{\beta}}^C,
\end{align}

\noindent holds up to total derivatives. 

The reprentation on the anti-chiral multiplet 

\begin{align}
	\Phi^A=
	\left( \begin{array}{@{}c@{}}
	A^{\ast} \\ \bar{\psi}^{\dot{\alpha}} \\ F^{\ast}
	\end{array} \right)
\end{align}

\noindent can be constructed in the same way as follows: The supersymmery transformation for the anti-chiral multiplet, 

\begin{align}
	& \delta_{\xi}A^{\ast}=\sqrt{2}(\bar{\xi}\bar{\psi}) , \nonumber \\
	& \delta_{\xi}\bar{\psi}^{\dot{\alpha}}
	=i\sqrt{2}(\bar{\sigma}^{\mu}\xi)^{\dot{\alpha}}
	\vec{\partial}_{\mu}A^{\ast}+\sqrt{2}\bar{\xi}^{\dot{\alpha}}F^{\ast}, 
	\nonumber \\
	& \delta_{\xi}F^{\ast}=i\sqrt{2}(\xi \sigma^{\mu}\vec{\partial}_{\mu}
	\bar{\psi}),
\end{align}

\noindent is realized by the supersymmetry generators represented on the anti-chiral multiplet in the matrix form, 

\begin{align}
	Q_{\beta}^A=
	\left( \begin{array}{ccc}
	0 & 0 & 0 \\
	i\sqrt{2}\epsilon_{\gamma \beta}
	\bar{\sigma}^{\mu \dot{\alpha}\gamma}\vec{\partial}_{\mu} & 0 & 0 \\
	0 & i\sqrt{2}\sigma_{\beta \dot{\alpha}}^{\mu}\vec{\partial}_{\mu} & 0
	\end{array} \right), \quad 
	\bar{Q}_{\dot{\beta}}^A=
	\left( \begin{array}{ccc}
	0 & \sqrt{2}\epsilon_{\dot{\beta}\dot{\alpha}} & 0 \\
	0 & 0 & -\sqrt{2}\delta_{\dot{\beta}}^{\> \dot{\alpha}} \\
	0 & 0 & 0
	\end{array} \right), 
\end{align}

\noindent and the inner product is defined through the matrix: 

\begin{align}
	I^A=\left( \begin{array}{ccc}
	-\frac{i}{2}\overleftrightarrow{\partial_0} & 0 & 0 \\
	0 & \frac{1}{2}\sigma^{0}_{\alpha \dot{\alpha}} & 0 \\
	0 & 0 & 0 \\
	\end{array} \right).
\end{align}

The same properties such as supersymmetry algebra (8) and hermiticity conditions (9) and (10) can be also confirmed for the anti-chiral multiplet $\Phi^A$ in the same way.

\vspace{0.5cm}

In passing to the 2nd quantized theory, the supersymmetry generator $\mathcal{Q}$ is expressed as 

\begin{align}
	\xi^{\beta}\mathcal{Q}_{\beta}+\bar{\xi}_{\dot{\beta}}
	\bar{\mathcal{Q}}^{\dot{\beta}}=\int d^3\vec{x} \> \Phi^{\dagger}I
	(\xi^{\beta}Q_{\beta}+\bar{\xi}_{\dot{\beta}}\bar{Q}^{\dot{\beta}})
	\Phi .
\end{align}

\section{On-shell Representation: 8-Component Representation}

So far, our formulation of supersymmetric quantum mechanics works well, but it contains some problems in the following reasons: 

\begin{enumerate}
	\item In principle, the operators of quamtum mechanics does not depend 
	on the time-derivative $\partial_0$ for the energy conserved system. 
	For example, the Hamiltonian $H$ in the Schr\"{o}dinger equation 
	$i\vec{\partial}_0\Phi =H\Phi$ should not contain $\partial_0$. 
	However, in our foregoing representation, supersymmetry generators 
	$\bar{Q}_{\dot{\beta}}^C$ and $Q_{\beta}^A$ contain $\partial_0$. 
	The time $t$ should be treated as a parameter for the model with the 
	time-independent Hamiltonian.
	\item The equation of motion for the scalar field $A$ is the 2nd order 
	differential equation with respect to time coordinate, so two fields 
	$A(t_0)$ and $\vec{\partial}_0A(t_0)$ are necessary as initial 
	condition at the time $t_0$. However $\Phi$ contains only $A$. It is 
	insufficient, since, in other words, the scalar field has the two 
	solutions, i.e. positive and negative energy ones and we need to 
	specify the sign of the energy by the field $\vec{\partial}_0A$.
	\item In connection with the above two reasons 1 and 2, the 
	Hamiltonian for scalar field $A$ is expressed as 
	$H=\pm \sqrt{\vec{\partial}_{i}^2+m^2}$, which are ill-defined 
	operators. And, as seen in the second equation of (3), the Majorana 
	fermion $\psi_{\alpha}$ and its conjugate $\bar{\psi}^{\dot{\alpha}}$ 
	enter in an equation together. Therefore we cannot define the 
	Hamiltonian acting on $\Phi^C$ and $\Phi^A$ separately. Of course, 
	this is not the case for the massless chiral and anti-chiral 
	multiplets.
\end{enumerate}

Then, in the next step, we would like to introduce the conjugate momentum $\Pi =\vec{\partial}_0A$ of $A$ and $\Pi^{\ast}=\vec{\partial}_0A^{\ast}$ of $A^{\ast}$, and rewrite $\Phi^C$ and $\Phi^A$ as 

\begin{align}
	\Phi^C=
	\left( \begin{array}{@{}c@{}}
	A \\ \Pi \\ \psi_{\alpha} \\ F
	\end{array} \right), \quad 
	\Phi^A=
	\left( \begin{array}{@{}c@{}}
	A^{\ast} \\ \Pi^{\ast} \\ \bar{\psi}^{\dot{\alpha}} \\ F^{\ast}, 
	\end{array} \right), 
\end{align}

\noindent respectively. We can express these as a 8-component field $\Phi$ as 

\begin{align}
	\Phi=\left( \begin{array}{@{}c@{}} \Phi^C \\ \Phi^A \end{array} \right)
	=\left( \begin{array}{@{}c@{}}
	A \\ \Pi \\ \psi_{\alpha} \\ F \\ A^{\ast} \\ \Pi^{\ast} \\ 
	\bar{\psi}^{\dot{\alpha}} \\ F^{\ast}
	\end{array} \right).
\end{align}

\noindent As a result, the supersymmetry generators and the matrices defining the inner product are expressed by $8\times 8$ matrices. By utilizing the equation of motions in order to eliminate the time derivative $\partial_0$, they read 

\begin{align}
	Q_{\beta}= & 
	\left( \begin{array}{cccccccc}
	0 & 0 & \sqrt{2}\delta_{\beta}^{\> \alpha} & 0 & 0 & 0 & 0 & 0 \\
	0 & 0 & -\sqrt{2}(\sigma^0\bar{\sigma}^i)_{\beta}^{\> \alpha}
	\partial_i & 0 & 0 & 0 & i\sqrt{2}m\sigma^0_{\beta \dot{\alpha}} & 0 \\
	0 & 0 & 0 & \sqrt{2}\epsilon_{\alpha \beta} & 0 & 0 & 0 & 0 \\
	0 & 0 & 0 & 0 & 0 & 0 & 0 & 0 \\
	0 & 0 & 0 & 0 & 0 & 0 & 0 & 0 \\
	0 & 0 & 0 & 0 & 0 & 0 & 0 & 0 \\
	0 & 0 & 0 & 0 & i\sqrt{2}\epsilon_{\gamma \beta}
	\bar{\sigma}^{i\dot{\alpha}\gamma}\partial_i & i\sqrt{2}
	\epsilon_{\gamma \beta}\bar{\sigma}^{0\dot{\alpha}\gamma} & 0 & 0 \\
	0 & 0 & -\sqrt{2}m\delta_{\beta}^{\> \alpha} & 0 & 0 & 0 & 0 & 0 
	\end{array} \right), \\
	\bar{Q}_{\dot{\beta}}= & 
	\left( \begin{array}{cccccccc}
	0 & 0 & 0 & 0 & 0 & 0 & 0 & 0 \\
	0 & 0 & 0 & 0 & 0 & 0 & 0 & 0 \\
	-i\sqrt{2}\sigma^i_{\alpha \dot{\beta}}\partial_i & 
	-i\sqrt{2}\sigma^0_{\alpha \dot{\beta}} & 0 & 0 & 0 & 0 & 0 & 0 \\
	0 & 0 & 0 & 0 & 0 & 0 & -\sqrt{2}m\epsilon_{\dot{\beta}\dot{\alpha}} & 
	0 \\
	0 & 0 & 0 & 0 & 0 & 0 & \sqrt{2}\epsilon_{\dot{\beta}\dot{\alpha}} & 
	0 \\
	0 & 0 & i\sqrt{2}m\epsilon_{\dot{\beta}\dot{\delta}}
	\bar{\sigma}^{0\dot{\delta}\alpha} & 0 & 0 & 0 & 0 & 0 \\
	0 & 0 & 0 & 0 & 0 & 0 & 0 & 
	-\sqrt{2}\delta_{\dot{\beta}}^{\> \dot{\alpha}} \\
	0 & 0 & 0 & 0 & 0 & 0 & 0 & 0 
	\end{array} \right), \\
	I= & 
	\left( \begin{array}{cccccccc}
	0 & \frac{i}{4} & 0 & 0 & 0 & 0 & 0 & 0 \\
	-\frac{i}{4} & 0 & 0 & 0 & 0 & 0 & 0 & 0 \\
	0 & 0 & -\frac{1}{4}\bar{\sigma}^{0\dot{\alpha}\alpha} & 0 & 0 & 0 & 
	0 & 0 \\
	0 & 0 & 0 & 0 & 0 & 0 & 0 & 0 \\
	0 & 0 & 0 & 0 & 0 & -\frac{i}{4} & 0 & 0 \\
	0 & 0 & 0 & 0 & \frac{i}{4} & 0 & 0 & 0 \\
	0 & 0 & 0 & 0 & 0 & 0 & \frac{1}{4}\sigma^0_{\alpha \dot{\alpha}} & 
	0 \\
	0 & 0 & 0 & 0 & 0 & 0 & 0 & 0 
	\end{array} \right).
\end{align}

\noindent Then we can represent the Hamiltonian in terms of $8\times 8$ matrix which is used for the Schr\"{o}dinger equation $i\partial_0\Phi =H\Phi$: 

\begin{align}
	i\partial_0
	\left( \begin{array}{@{}c@{}}
	A \\ \Pi \\ \psi_{\alpha} \\ F \\ A^{\ast} \\ \Pi^{\ast} \\ 
	\bar{\psi}^{\dot{\alpha}} \\ F^{\ast}
	\end{array} \right)=
	\left( \begin{array}{cccccccc}
	0 & i & 0 & 0 & 0 & 0 & 0 & 0 \\
	i(\partial_i^2-m^2) & 0 & 0 & 0 & 0 & 0 & 0 & 0 \\
	0 & 0 & -i(\sigma^0\bar{\sigma}^i)_{\alpha}^{\> \beta}\partial_i & 
	0 & 0 & 0 & -m\sigma^0_{\alpha \dot{\beta}} & 0 \\
	0 & 0 & 0 & 0 & 0 & 0 & 0 & 0 \\
	0 & 0 & 0 & 0 & 0 & i & 0 & 0 \\
	0 & 0 & 0 & 0 & i(\partial_i^2-m^2) & 0 & 0 & 0 \\
	0 & 0 & -m\bar{\sigma}^{0\dot{\alpha}\beta} & 0 & 0 & 0 & 
	-i(\bar{\sigma}^0\sigma^i)_{\> \dot{\beta}}^{\dot{\alpha}}\partial_i & 
	0 \\
	0 & 0 & 0 & 0 & 0 & 0 & 0 & 0 
	\end{array} \right)
	\left( \begin{array}{@{}c@{}}
	A \\ \Pi \\ \psi_{\beta} \\ F \\ A^{\ast} \\ \Pi^{\ast} \\ 
	\bar{\psi}^{\dot{\beta}} \\ F^{\ast}
	\end{array} \right).
\end{align}

\noindent We note that, as stated above, $Q_{\beta},\bar{Q}_{\dot{\beta}}$ and $H$ are reducible to chiral and anti-chiral multiplets, i.e. $4\times 4$ matrices, for the massless case $m=0$.

\vspace{0.5cm}

To close this section, we note that the above argument can easily be repeated in a superspace formalism, because each entry of the above matrix representation (21) is nothing but the degree of superspace coordinate $\theta$. As is well known, instead of (5), we may use the supersymmetry generators in a superspace formalism in terms of chiral superfield in the following form, 

\begin{align}
	& Q_{\alpha}=\frac{\partial}{\partial \theta^{\alpha}}
	-i\sigma_{\alpha \dot{\alpha}}^{\mu}\bar{\theta}^{\dot{\alpha}}
	\partial_{\mu}, \quad \bar{Q}_{\dot{\alpha}}=	
	-\frac{\partial}{\partial \bar{\theta}^{\dot{\alpha}}}
	+i\theta^{\alpha}\sigma_{\alpha \dot{\alpha}}^{\mu}\partial_{\mu}.
\end{align}

\noindent By making use of the Lorentz invariant inner product

\begin{align}
	\langle \Phi |\Phi \rangle =\int d^3\vec{x} d^2\theta d^2\bar{\theta}
	\> \Phi^{\dagger}\theta \sigma^0 \bar{\theta}\Phi ,
\end{align}

\noindent where we used $\theta\sigma^0\bar{\theta}$ instead of the matrix I in equations (6), (7) and (14). In this way we go on the argument in the superspace formalism exactly parallel to our previous formalism where the matrix representation is used. However, in this superspace formalism, it is difficult to treat the conjugate momenta $\Pi$ and $\Pi^{\ast}$, since we are not sure what degrees of $\theta$ and $\bar{\theta}$ they come in. Therefore the matrix formulation constructed so far is very preferable.

\section{Conclusions}

\vspace{0.5cm}

In this paper, we presented an attempt to formulate the supersymmetric relativistic quantum mechanics to be understood in the sense that we considered a single particle that had it as an internal degree of freedom to be of different species deviating even with respect to the spin being integer or half integer. A symmetry transformation given by our Q-matrices formula (18) and (19) transforming between the states of different half-integerness can with great right be called a supersymmetry generator although strictly speaking there is of course no sense to the notion of Fermions and Bosons as long as we consider only ONE particle. We found that our representation of supersymmetry generators in the matrix form is properly constructed only when we utilize the equations of motion (Schr\"{o}dinger equations) and can be interpreted as the supersymmetry generators in the second quantization. Therefore, we may conclude that our supersymmetry generators in the matrix form are useful to formulate the relativistically supersymmetric quantum mechanics.

However there is the unsolved problem to the authors: It is to derive the classical action which should lead to the supersymmetric quantum mechanics. It is still unknown how to obtain this classical action. On the other hand, in the usual theories, as we know, the Schr\"odinger equation (3) that is derived from the classical action describing world line, is obtained by replacing the coordinates of the space-time by operators. However, we have not succeeded so far to obtain the classical action in supersymmetric case.

\vspace{1.5cm}

\noindent \underline{ Acknowledgement }

\vspace{0.25cm}

This work is supported by Grants-in-Aid for Scientific Research on Priority Areas, Number of Areas 763, ``Dynamics of strings and Fields", from the Ministry of Education of Culture, Sports, Science and Technology, Japan.

\vspace{2.5cm}
\noindent{\bf \Large Appendix}
\vspace{0.5cm}

\noindent The Minkowski metric is $\eta^{\mu \nu}=(-1,+1,+1,+1)$ and index $i$ represents the space 3-vector. We may summarize in the following our notations used in the present paper. 

\vspace{0.5cm}

\noindent Spinors: 

\begin{align*}
	\psi_{\alpha}
	=\left(\begin{array}{@{}c@{}} \psi_1 \\ \psi_2 \end{array}\right), 
	\quad 
	\bar{\psi}^{\dot{\alpha}}
	=\left(\begin{array}{@{}c@{}}
	\bar{\psi}^{\dot{1}} \\ \bar{\psi}^{\dot{2}}
	\end{array}\right).
\end{align*}

\noindent Convensions: 

\begin{align*}
	& \epsilon_{21}=\epsilon^{12}=\epsilon_{\dot{2}\dot{1}}
	=\epsilon^{\dot{1}\dot{2}}=1, \\
	& \psi^{\alpha}=\epsilon^{\alpha \beta}\psi_{\beta}, \quad 
	\psi_{\alpha}=\epsilon_{\alpha \beta}\psi^{\beta}, \\
	& \psi_{\dot{\alpha}}=\epsilon_{\dot{\alpha}\dot{\beta}}
	\psi^{\dot{\beta}}, \quad 
	\psi^{\dot{\alpha}}=\epsilon^{\dot{\alpha}\dot{\beta}}
	\psi_{\dot{\beta}}, \\
	& (\psi \xi )=\psi^{\alpha}\xi_{\alpha}, \quad 
	(\bar{\psi}\bar{\xi})
	=\bar{\psi}_{\dot{\alpha}}\bar{\xi}^{\dot{\alpha}}.
\end{align*}

\noindent Sigma matrices: 

\begin{align*}
	& \sigma^0=
	\left(\begin{array}{cc} -1 & 0 \\ 0 & -1 \end{array}\right), \quad 
	\sigma^1=
	\left(\begin{array}{cc} 0 & 1 \\ 1 & 0 \end{array}\right), \\
	& \sigma^2=
	\left(\begin{array}{cc} 0 & -i \\ i & 0 \end{array}\right), \quad 
	\sigma^3=
	\left(\begin{array}{cc} 1 & 0 \\ 0 & -1 \end{array}\right), \\
	& \bar{\sigma}^{\mu \dot{\alpha}\alpha}
	=\epsilon^{\dot{\alpha}\dot{\beta}}\epsilon^{\alpha \beta}
	\sigma^{\mu}_{\beta \dot{\beta}}.
\end{align*}

\noindent Fierz transformation: 

\begin{align*}
	(\psi \phi )\bar{\xi}_{\dot{\alpha}}
	=-\frac{1}{2}(\phi \sigma^{\mu}\bar{\xi})
	(\psi \sigma_{\mu})_{\dot{\alpha}}.
\end{align*}


\end{document}